\documentstyle[12pt,aaspp4,flushrt]{article}
\def\approxlt{\lower.2em\hbox{$\buildrel < \over \sim$}}

\def\itb{{\it B\/}}
\def\iti{{\it I\/}}
\def\itk{{\it K\/}}

\def\ang{\thinspace{\rm \AA}}
\def\msun{~\rm M_{\odot}}

\def\mdot{$\dot{M}\/$}
\def\kms{~{\rm km\ sec}^{-1}}

\def\ha{\ifmmode {{\rm H}\alpha}
        \else {H$\alpha$}\fi}
\def\lstar{\ifmmode {L_{\ast}}
        \else {$L_{\ast}$}\fi}
\def\mkstar{M_{\rm K\ast}}
\def\secpoint{^{\prime\prime}\kern-2.1mm .\kern+.6mm}
\def\oii{[O\thinspace{\sc{ii}}]}
\def\nii{[N\thinspace{\sc{ii}}]}

\def\phistar{\phi_{\ast}}
\def\fnu{$F_{\nu}$}

\def\hnought{{\rm H}_0}
\def\qnought{\ifmmode q_0
    \else $q_0$\fi}

\def\ten#1{\ifmmode 10^{#1}
    \else $10^{#1}$\fi}
\newcount\hours
 \newcount\minutes
  \def\SetTime{\hours=\time
         \global\divide\hours by 60
         \minutes=\hours
         \multiply\minutes by 60
         \advance\minutes by-\time
         \global\multiply\minutes by-1 }
 \SetTime
 \def\now{\number\hours:\ifnum\minutes<10 0\fi\number\minutes}

\begin{document}
\slugcomment{To appear in {\it The Astrophysical Journal (Letters)}}

\title{The Evolution of the Distribution of Star Formation Rates in Galaxies}
\author{Lennox L. Cowie,\altaffilmark{1,2} Esther M. Hu,\altaffilmark{1,2}
  Antoinette Songaila\altaffilmark{1}}
\affil{Institute for Astronomy, University of Hawaii, 2680 Woodlawn Dr.,
  Honolulu, HI 96822\\
  cowie@ifa.hawaii.edu, hu@ifa.hawaii.edu, acowie@ifa.hawaii.edu}
\and
\author{Eiichi Egami\altaffilmark{2}}
\affil{Max-Planck-Institut f\"{u}r extraterrestrische Physik, 
  85740 Garching, GERMANY\\
  egami@mpe-garching.mpg.de}

\altaffiltext{1}{Visiting Astronomer, W. M. Keck Observatory, jointly
  operated by the California Institute of Technology and the University of
  California.}
\altaffiltext{2}{Visiting Astronomer, United Kingdom Infrared Telescope,
  operated by the Joint Astronomy Centre on behalf of the U. K. Particle
  Physics and Astronomy Research Council.}


\begin{abstract}
A large deep and nearly complete $B<24.5$ redshift sample is used to
measure the change in distribution function of the stellar mass production 
rate in individual galaxies with redshift.  The evolution of the star formation
rate distribution with redshift is interpreted in terms of the history of
spiral galaxy formation, with the disk component modelled as a single
evolving entity, and the characteristic timescales, luminosities, and
epochs varying according to galaxy type.  The more massive forming
galaxies seen at $z=1\to3$ are identified as earlier type spirals, whose
star formation rates are initially high and then decline rapidly at $z<1$,
while for later type spirals and smaller mass irregulars the mass
formation rates at $z<1$ are lower, and the formation process persists to
redshifts much closer to the present epoch.  We find that these models can
be consistent with the data and fit well into a broad picture of other
recent results if \qnought=0.02 and many of the disks begin their growth
at $z\ll3$, but that they predict too many bright star formers at high $z$
in flat universes.

\end{abstract}


\section{Introduction}

Disks of spiral galaxies are extremely fragile and must have formed
primarily by conversion of gas to stars in an essentially invariant galaxy
potential with only a small accretion of external stars (e.g.,
\markcite{toth92}Toth \& Ostriker 1992).  Thus, the stellar history of
this component may be approximately modelled as the evolution of a single
entity, which greatly simplifies predicting the luminosity evolution.  In
the present Letter we trace the evolution of the distribution of star
formation rates in galaxies with redshift to determine whether it is
consistent with this type of pure luminosity evolution.  The problem is
relatively straightforward because beyond $z\sim 0.3$ the observed
\itb\ band corresponds to ultraviolet light ($\lambda\ll3500$\ang), which
in all but the oldest galaxies is a measure of the ongoing rate of massive
star formation.  The \itb\ luminosity is then directly proportional to the
rate of metal production in the individual galaxy, while the
contribution of the ensemble of galaxies to the extragalactic background
light measures the overall metal density production in the universe
(\markcite{cow88}Cowie 1988; \markcite{son90}Songaila, Cowie, \& Lilly
1990; \markcite{lil96}Lilly et al.\ 1996; \markcite{mad96}Madau et
al.\ 1996).  Viewed this way, deep $B$-band redshift samples measure the
distribution function of the stellar mass production rate in galaxies,
which we will refer to as the \mdot\ function, in contrast to infrared
samples, which for all but the youngest galaxies measure the distribution
of the mass in galaxies.

In order to probe star formation rates (SFR) in individual galaxies to
high redshifts, large and extremely deep $B$-band redshift samples are
required.  Here we use a large, 95\% complete sample of galaxies to $B$ =
24.5 with redshifts to $z=2.2$, to measure the evolution of the
\mdot\ function with redshift.  We find that smooth single-entity
evolution models can work if \qnought=0.02 and many disks begin to form
only at later epochs, but predict too many bright star formers at high $z$
in flat universes.

\section{Data}\label{sec:data}

\markcite{cow96}Cowie et al.\ (1996) described LRIS observations of a
$B$ = 24.5 galaxy sample in the Hawaii fields SSA13 and SSA22 (26.2
arcmin$^2$).  However, even the more intensively studied SSA22 field was
only 84\% complete at $B=24-24.5$, with 13 $B\leq24.5$ galaxies
unidentified.  Since extremely high completeness is required to determine
the \mdot\ function and the unidentified objects may correspond to higher
redshift objects in the sample, we made intensive efforts to complete
identifications of the remaining $B\leq 24.5$ galaxies in SSA22 using
deeper optical (3500--10000\ang) observations and near infrared
spectroscopy.  The improved S/N and additional spectral features yield
robust identifications for all but four sample objects.

The SSA22 $B<24.5$ sample is used with a nearly complete $B<24$ sample in
SSA13.  In the $B < 24$ sample covering 26 arcmin$^2$ there are 156
objects (124 galaxies) of which all but 4 are identified (97\% complete),
while in the $B = 24-24.5$ range covering 13 arcmin$^2$ there are 47
objects (42 galaxies) of which all but 2 are identified (95\% complete).
The six unidentified galaxies have extremely blue colors but no strong
\oii, with most likely redshifts $1.6 < z < 2$, where the \oii\ line has
moved from the observed optical while many stronger UV features have not
yet entered the blue wavelength range.  However, some objects may lie at
somewhat higher redshifts.

\begin{figure}
\plotone{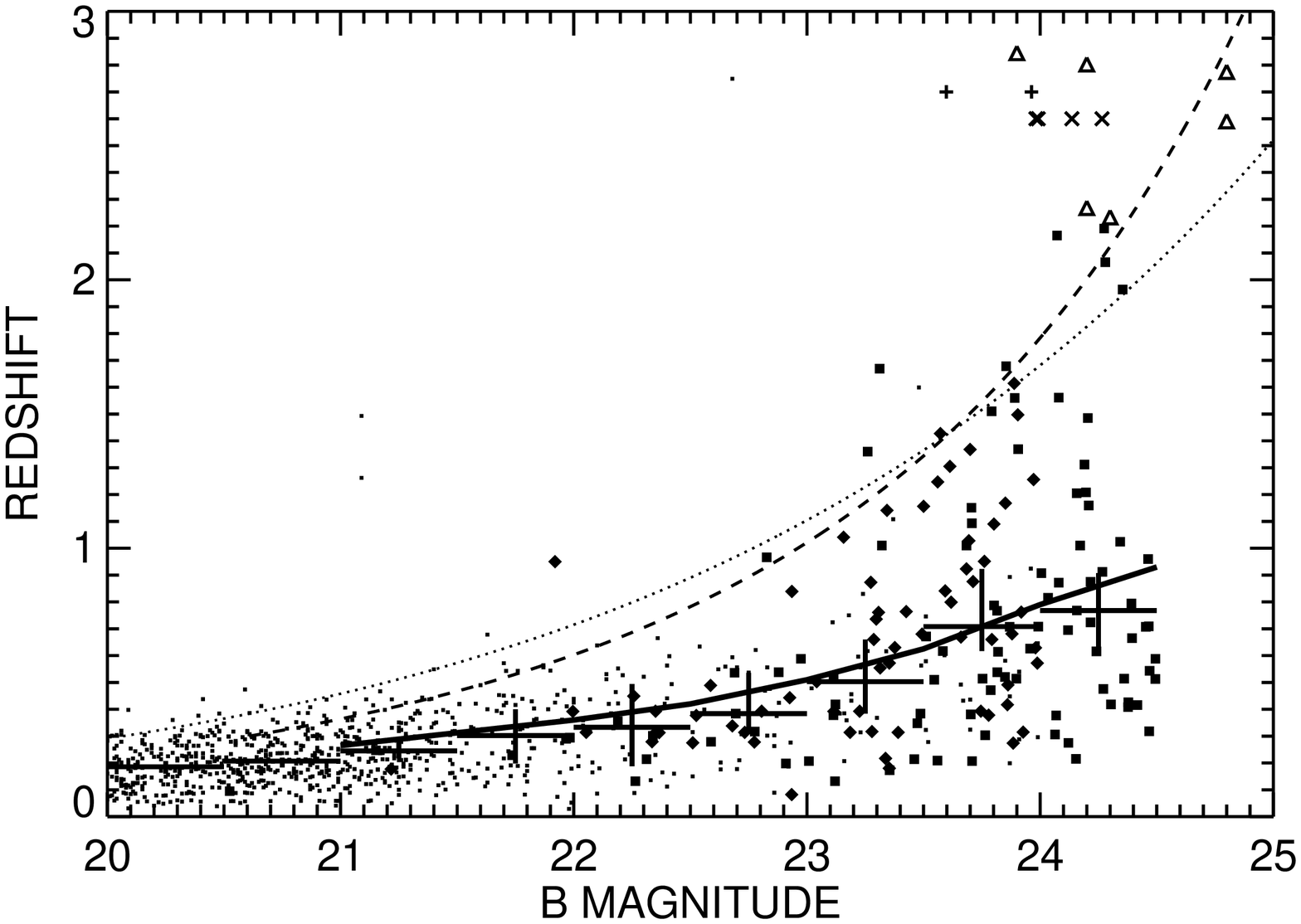}
\caption{\label{fig:1}The redshift-$B$ magnitude distribution for SSA22 $B<24.5$
(solid boxes) and SSA13 $B<24$ (diamonds).  Unidentified
objects are shown as crosses at nominal redshifts of $z=2.6$ (SSA22) and
$z=2.7$ (SSA13).  The very small symbols show the large Autofib sample of
\protect{\markcite{ell96}}Ellis et al.\ (1996).  The few points lying
outside of the well-populated regions are AGN.  The large crosses show the
median redshifts computed in half-magnitude intervals with $\pm1\sigma$
error bars computed using the median sign method.  (The Autofib data are
included at $B<23$ only.)\ \ These are very well described by the
no-luminosity-evolution model of \protect{\markcite{gla95a}}Glazebrook et
al.\ (1995a) (solid line).  The dashed (\qnought=0.5)
and dotted (\qnought=0.02) lines show the \itb-magnitude vs.\ redshift
relation for a flat $F_{\nu}$ galaxy which roughly matches the upper
envelope of the observed objects in the $z=1-2$ range.  We also show (open
triangles) the six $B<25$, $z>1.6$ galaxies which have been identified in
the Hubble Deep Field to date (\protect{\markcite{coh96}}Cohen et al.\ 1996;
\protect{\markcite{stei96b}}Steidel et al.\ 1996b;
\protect{\markcite{low97}}Lowenthal et al.\ 1997).  This rather
heterogeneously selected sample shows rough consistency with the upper
envelope of the $B-z$ relation seen in the magnitude-selected sample, 
as do other galaxies selected with the
ultraviolet break color techniques (\protect{\markcite{stei96a}}Steidel et
al.\ 1996a), though they may also contain a small number of more luminous
galaxies.
}
\end{figure}

The redshift-magnitude relation and median redshifts are shown in
Figure~\ref{fig:1}, along with the Autofib sample of \markcite{ell96}Ellis
et al.\ (1996).  As at brighter magnitudes, the median \itb\ redshifts,
computed from the combination of the Autofib data plus the present sample,
are well described by a model (solid line) in which the luminosity
function remains invariant (\markcite{bro88}Broadhurst, Ellis, \& Shanks
1988; \markcite{col90}Colless et al.\ 1990; \markcite{cow91}Cowie,
Songaila, \& Hu 1991; \markcite{gla95a}Glazebrook et al.\ 1995a;
\markcite{cow96}Cowie et al.\ 1996), though this reflects a much more
complex underlying situation.  Dashed and dotted lines show expected
magnitudes for a flat \fnu\ galaxy with $M_{AB} = -19.8 + 5\,{\rm
log}\,h_{65}$ for \qnought=0.5 (dashed) and $M_{AB} = -20.5 + 5\,{\rm
log}\,h_{65}$ for \qnought=0.02 (dotted), which roughly map the upper
envelope of the magnitude-redshift relation at $1<z<2$.  As we shall see
below these correspond to star formation rates of roughly
$9\,h_{65}^{-2}\ \msun$ yr$^{-1}$ (\qnought=0.5) and $16\,h_{65}^{-2}\
\msun$ yr$^{-1}$ (\qnought=0.02).  As  \markcite{cow96}Cowie et
al.\ (1996) emphasized, there are few galaxies with such high star
formation rates at $z\ll 1$, and the envelope of the observed distribution
lies well to the right of the curves at these redshifts, except for a few
luminous AGN present in the samples.  The brighter $z=2-3$ galaxies lie to
the left of the line, with mass formation rates up to a factor of 2.5
higher.  (The Autofib and HDF data are not used in the remainder of the
paper in order to avoid completeness problems.)

\section{Interpretation}\label{sec:interpret}

The present $B < 24.5$ sample provides two independent measurements of the 
massive star formation rate.  The first is the rest-frame far UV ($\lambda
\ll 3500$\ang) light, which for galaxies with significant ongoing star
formation is a direct measure of the massive star formation rate, and
hence of the metal production rate (\markcite{cow88}Cowie 1988;
\markcite{son90}Songaila et al.\ 1990).  Translating this to a total 
star formation rate requires an uncertain assumption about the slope
of the IMF.  For a Salpeter IMF extending over the range $0.1\to125\msun$ 
we may calibrate from the \markcite{bru96}Bruzual \& Charlot
(1996) models to obtain, in the absence of internal extinction,
\begin{equation}
        \dot{M} = 100\,{\msun}\,{\rm yr}^{-1} \left ({L\,(2500\ang)}\over
        {6.7\times\ten{29}\ {\rm ergs\ s}^{-1}\ {\rm Hz}^{-1}} \right )\ ,
	\label{eq:mdot}
\end{equation}
where a rest-frame wavelength of 2500\ang\ is chosen to 
minimize extrapolations from the $B$ band in lower redshift galaxies
while also minimizing the slight contamination by older stars which is
present at slightly longer wavelengths.  Equation~(\ref{eq:mdot}) may be 
derived from extremely simple physical arguments and is quite model
independent except for the assumed IMF (\markcite{cow88}Cowie 1988).
2500\ang\ corresponds to $B$ at $z=0.8$ and $I$ at $z=2.4$, and may be 
interpolated at intermediate redshifts using approximate galaxy types 
inferred from the (\itb, \iti, \itk) galaxy colors, so that the 
2500\ang\ luminosity can be relatively accurately determined at least at 
the higher redshifts.  The massive star formation rate can also be obtained 
from the line luminosities in the galaxies, which measure the production 
rate of ionizing photons.  The most direct diagnostic is $L(\ha)$, which 
can be measured in these galaxies from $z=0 \to 
2.3$.  For most of the galaxies, however, \ha\ is not measured, 
and we must draw on the secondary indicator \oii\ 3727\ang, whose 
equivalent width is closely related to that of \ha\ by the relationship 
W$_{\lambda}$(\ha+\nii) = 2.3 W$_{\lambda}$(\oii) in a wide variety of 
local and moderate redshift galaxies (\markcite{ken92}Kennicutt 1992; 
\markcite{son94}Songaila et al.\ 1994).  For local blue 
irregular galaxies (\markcite{gal89}Gallagher, Bushouse, \& Hunter 1989) 
\begin{equation}
   \dot{M} = 100\,{\msun}\,{\rm yr}^{-1} \left ({L\,({\rm [O\thinspace{II}])}}
        \over {\ten{43}\ {\rm ergs\ s}^{-1}} \right )\ ,
        \label{eq:oii}
\end{equation}
but we derive an essentially identical calibration for the present galaxies
by comparison with the more direct UV luminosity calibration, suggesting
that this calibration is reasonable for most rapidly star-forming galaxies
irrespective of mass and redshift.  The \oii\ luminosity provides an
independent check on the UV luminosity which is particularly valuable at
the low redshift end (\markcite{cow95}Cowie, Hu, \& Songaila 1995).

Internal extinction in the galaxies is a serious concern for
both diagnostics, since escaping UV photons can be reduced
significantly by this effect, while the internal production of \ha\ and
\oii\ can be reduced if ionizing photons are extinguished.  However, at
least for the higher redshift blue galaxies there appears to be remarkably
little extinction and this effect appears weak.  The great bulk of the
$z>0.8$ galaxies are extremely flat in the rest-frame $2500\ang-21000\ang$
color, with a median value of 0.9 in the $AB\/$ system, and have high
rest-frame \oii\ equivalent widths (with a median of 40\ang).  Since the
intrinsic spectrum is expected to be flat or falling with increasing
frequency, this places very severe constraints on the amount of dust.
For a \markcite{sea79}Seaton (1979) law $A_{2500}
\approx\ E(2500\ang-10000\ang) \approx\ 7\,E(B-V)$.  Even if 
\fnu\ is intrinsically flat, $A_{2500} = 0.9$ on average, and at most half
the light is lost to extinction. Thus, the maximum uncertainty
introduced by not including an extinction correction is probably a factor
of 2.  To allow for the extinction we therefore assume that the
luminosities in equations (\ref{eq:mdot}) and (\ref{eq:oii}) are reduced by
a factor of 1.5 for a given mass formation rate.

\begin{figure}
\figurenum{2}
\vbox to2.20in{\rule{0pt}{2.20in}}
\includegraphics{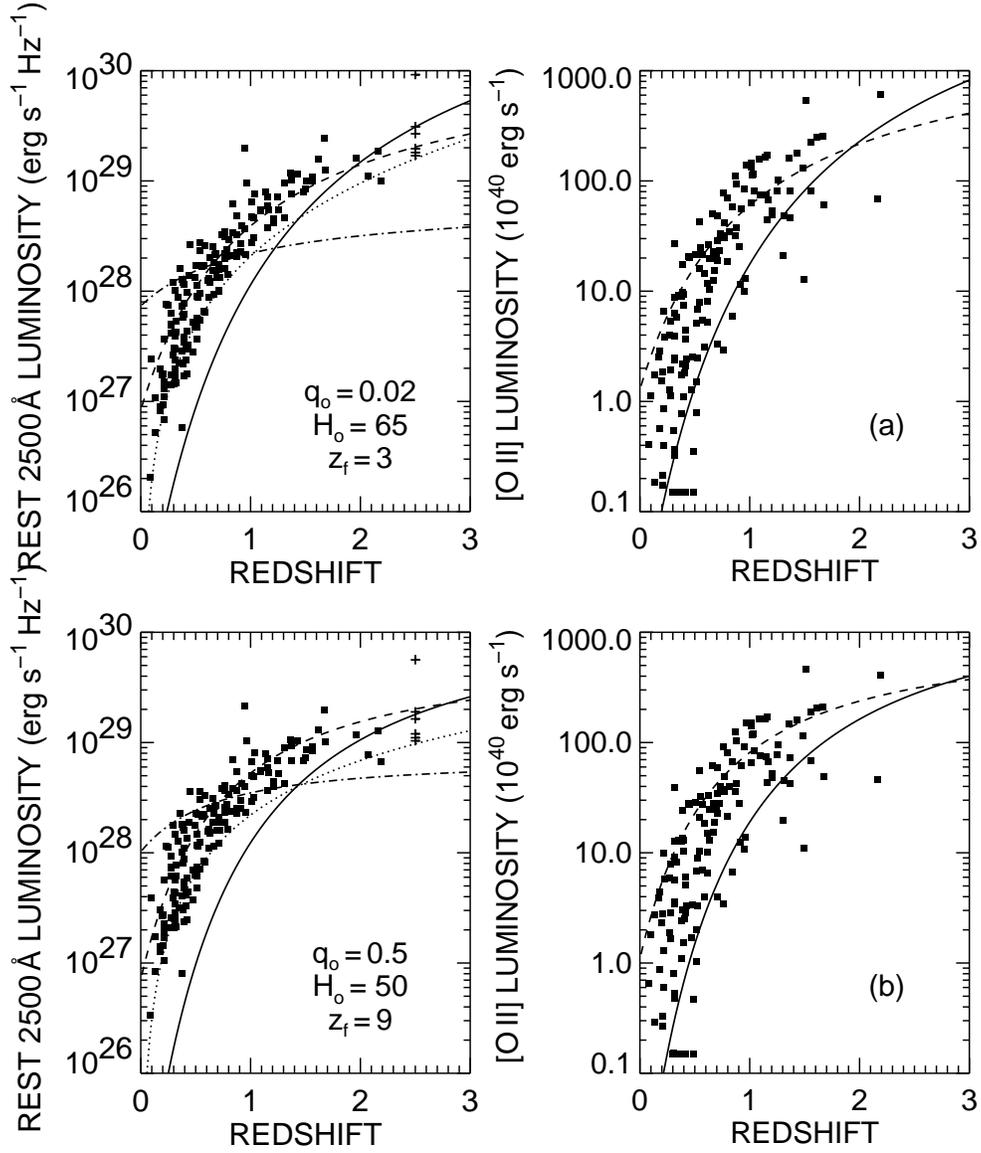}
\vskip-0.1in
\caption{\label{fig:2}The rest frame $2500\ang$ luminosity (left panels) and
\oii\ luminosity (right panels) computed for two geometries ---
\qnought=0.02 and $\hnought=65\kms$ Mpc$^{-1}$ (a) and \qnought=0.5 and
$\hnought=50\kms$ Mpc$^{-1}$ (b).  In the left-hand panels the dotted line
shows the selection limit for a $B=24.5$ galaxy with a flat $F_{\nu}$
spectrum.  Unidentified objects are shown as crosses at a nominal redshift
of 2.5.  Models are shown for exponentially decaying star formation rates
which would form a $1.4\times\ten{11}\ h_{65}^{-2}\ \msun$ galaxy with
$\tau = 1$ Gyr (solid line), and $\tau = 2$ Gyr (dashed line), or a
$7\times\ten{10}\ h_{65}^{-2}\ \msun$ galaxy with $\tau = 7$ Gyr (dash-dot
line).  For the open geometry the galaxy evolution is started at $z_f=3$
and for the flat geometry at $z_f=9$.
}
\end{figure}

The computed rest-frame 2500 \AA\ and \oii\ luminosities are shown versus
redshift in Figure~\ref{fig:2}. For \qnought=0.02 and $\hnought=65\kms$
Mpc$^{-1}$ (Fig.~\ref{fig:2}a) the upper envelope of the 2500$\,$\AA\
luminosity at $z>1$ lies at approximately $\ten{29}\ h_{65}^{-2}$ ergs
s$^{-1}$ Hz$^{-1}$ and just above $\ten{42}\ h_{65}^{-2}$ ergs s$^{-1}$
for \oii, which correspond to mass formation rates in the $15-30\
h_{65}^{-2}\ \msun$ yr$^{-1}$  range.  No objects with much higher mass
formation rates are seen.  At lower redshifts the upper envelope falls
rapidly.  For \qnought=0.5 we have used $\hnought=50\kms$ Mpc$^{-1}$
(Fig.~\ref{fig:2}b) to provide an acceptable age for the universe.  With
this smaller Hubble constant the maximum luminosities and mass formation
rates are similar to those in the \qnought=0.02 case.

\begin{figure}
\plotone{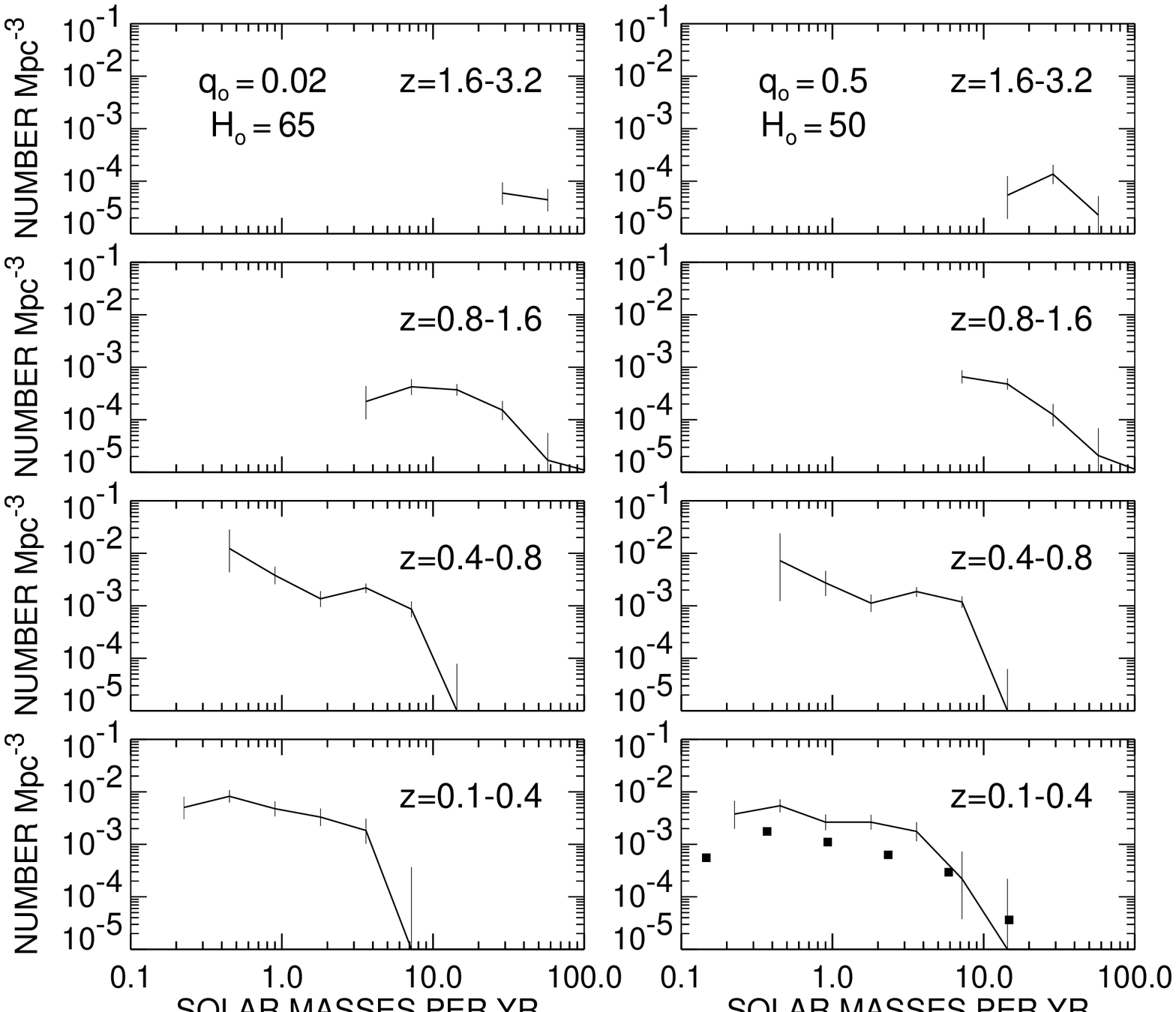}
\vskip0.1in
\caption{\label{fig:3}The distribution function of the star formation rates 
as a function of redshift interval for \qnought=0.02 (left panels) and
\qnought=0.5 (right panels).  The vertical axis shows the number of
galaxies per Mpc$^3$ per logarythmic bins of 0.3. The errors are 
$\pm1\sigma$ based on the number of objects in each bin.  The filled squares 
show the local ($z\approxlt0.045$)
SFR of \protect{\markcite{gal95}Gallego et al.\ (1995)}, plotted for the
Salpeter IMF assumed for the present calculations.  (The \ha\ luminosity
corresponding to a star formation rate of 1 solar mass per year is a factor
of 3 times higher than for the Scalo IMF, less rich in massive stars,
used in the Gallego et al.\ paper.)\ Error bars ($\pm1\sigma$)
for these points, based on the number of objects per bin,
are comparable to the symbol size; for the endmost points, they are roughly
twice the symbol size.  The agreement in shape between these two curves is good,
with the SFR for the $z=0.1-0.4$ range approximately a factor of two higher
than for the \protect{\markcite{gal95}Gallego et al.} sample. Differences
between the two curves are likely to arise due to differences in each sample's 
completeness in identifying star-forming galaxies, as well as a possible 
deficiency in local galaxies (\protect{\markcite{hua97}Huang et al.\ 1997}).  
}
\end{figure}

The distribution of $\dot M$ in various redshift bins, computed using the
$1/V$ method, is shown for open and flat geometries in Figure~\ref{fig:3}.
The star formation rate may be directly compared with the local star
formation rate determined by \markcite{gal95}Gallego et al.\ (1995) from
their \ha\ survey, which covered redshifts $z\approxlt 0.045$, for a 470
deg$^2$ sample with EW(\ha+\nii)$ > 10$\AA.  This is shown for $z=0.1-0.4$
in the bottom right panel of Figure~\ref{fig:3}.  The data points lie
roughly a factor of two higher than the \markcite{gal95}Gallego et
al.\ points, although both data sets are fully consistent in slope within
the respective 2$\sigma$ errors and uncertainties.  Interestingly, this
same difference betwen the number density of local vs.\ distant galaxies is
seen in the $K$-band count analysis of \markcite{hua97}Huang et
al.\ (1997), who suggested that there is a local deficiency of galaxies by
a factor of 2 on scale sizes out to $\sim300\ h^{-1}$ Mpc.  Applying such a
factor of 2 correction to the local SFR normalization would then directly
match the data at $z=0.1-0.4$.

There is a significant evolution in the maximum mass formation rates with 
redshift for both open and flat geometries, in the sense that galaxies
with much higher formation rates are seen at the higher redshifts (e.g.,
\markcite{cow96}Cowie et al.\ 1996).  For $z=0.8-1.6$ there are 27
galaxies with $\dot M > 10 \msun$ yr$^{-1}$ while none are seen at
$z<0.8$.  Even allowing for the relative volumes the probability of their
being drawn from the same distribution function is only $6\times\ten{-3}$
for \qnought=0.02 and $3\times\ten{-3}$ for \qnought=0.5.  However, the
volume density of the rapid star formers is not high.  For \qnought=0.02
the volume density of objects with $\dot M > 10\,h_{65}^{-2} \msun$
yr$^{-1}$ is $5.4\times\ten{-4}\,h_{65}^{-3}$ Mpc$^{-3}$ at $z=0.8-1.6$
and $1.0\times\ten{-4}\,h_{65}^{-3}$ Mpc$^{-3}$ for $z=1.6-3.2$, while for
\qnought=0.5 and $\dot M > 10\,h_{50}^{-2} \msun$ yr$^{-1}$ the densities
are $6.2\times\ten{-4}\,h_{50}^{-3}$ Mpc$^{-3}$ at $z=0.8-1.6$ and
$2.2\times\ten{-4}\,h_{50}^{-3}$ Mpc$^{-3}$ for $z=1.6-3.2$.

\section{Discussion}\label{sec:discuss}

In order to connect these results to the present epoch we need the mass
distribution and universal mass density of the local galaxy sample, and
these are best derived from the \itk-band observations.  
There is still some disagreement in the derived values of the local \itk-band 
luminosity function, and we adopt here the average of
the samples of \markcite{mob93}Mobasher, Sharples, \& Ellis (1993),
\markcite{gla95b}Glazebrook et al.\ (1995b), and \markcite{cow96}Cowie et
al.\ (1996), which corresponds to a Schechter function $\mkstar =
-24.2$\ ($+5 \log_{10}\,h_{65}$), $\alpha=-1.1$, and $\phistar = 2 \times
10^{-3}\,h_{65}^3~{\rm Mpc}^{-3}$.  The disagreement in $\mkstar$ is
$\sim0.5$ mags.  Most of the light density of the universe is
contained in objects near \lstar, with 50\% of the light density in objects
brighter than $\langle L\rangle = 0.7 \lstar$.  Most of these brighter
galaxies are earlier than Sb based on their \itb, \iti, \itk\ colors
(\markcite{hua97}Huang et al.\ 1997), and for consistency we have converted
from \itk\ magnitudes to masses using the \markcite{bru96}Bruzual \&
Charlot (1996) models, with \fnu\ $= 2.5\times\ten{29}$
ergs s$^{-1}$ Hz$^{-1}$ at 21000\ang\ corresponding to a stellar mass of
$\ten{11}\ \msun$, whence \lstar\ corresponds to
$1.4\times\ten{11}\ h_{65}^{-2}\ \msun$.  The local stellar mass density of
the universe is then $2.8\times\ten{8}\ h_{65}^{-2}\ \msun$ Mpc$^{-3}$.

The present-day spectra of the spiral galaxies are well represented by
evolutionary models with exponentially decaying star formation rates.
\markcite{bru93}Bruzual \& Charlot (1993) fit an Sb galaxy with a
$\tau_{SFR} = 2$ Gyr and an age of 8 Gyr, and an Sc galaxy with a
$\tau_{SFR} = 7$ Gyr and an age of 12 Gyr.  In Figure~\ref{fig:2} we show
the predicted luminosities for such exponentially evolving galaxies, with
masses corresponding to \lstar\ for $\tau_{SFR} = 1$ Gyr and $\tau_{SFR} =
2$ Gyr, and to 0.5 \lstar\ for $\tau_{SFR} = 7$ Gyr, for comparison with
the observed luminosities.  Because both $\lstar$ and the observed galaxy
luminosities scale as $h^{-2}$, the curves can only be adjusted via the
x-axis conversion from redshift to time and by the choice of the geometry,
so the absolute agreement in the normalization of observed and
predicted luminosities with redshift shows impressive
consistency.  For \qnought=0.02 we can fit the envelope with an
$\hnought=65\kms$ Mpc$^{-1}$ and a late formation epoch $z_f=3$ model, but
for \qnought=0.5 we need a low Hubble constant ($\hnought\approxlt50\kms$
Mpc$^{-1}$) and an early formation epoch ($z_f\sim10$).  Basically, 
to match the fall of the maximum luminosity with redshift at
$z<1$, we require galaxy ages of very roughly 4 Gyr at
$z=1$.  The expected luminosity envelope is better matched for the
$\qnought=0.02$ case, while for \qnought=0.5 there are fewer high
luminosity galaxies at high redshift to form the more massive spiral
galaxies.  Later type galaxies, such as the 0.5 \lstar\ Sc shown by the
dotted line in Figure~\ref{fig:2}, are also consistent with the
observations provided they are not too massive, so that their luminosities
are not too high at the lower redshifts.  A requirement of the models is
therefore that the later type galaxies should have lower \itk\ luminosities
than the earlier types, with Scs being at least a magnitude fainter than
Sbs, and Sds being considerably fainter still.  The current \itk-band data
are not yet adequate to robustly investigate type dependence in the
luminosity function, but this prediction should shortly be testable.

We can now quantify the problem with the \qnought=0.5 geometry:  Predicted
luminosities of the early stages of near-\lstar\ spirals are well above the
detection limit (Fig.~\ref{fig:2}b), at least for galaxies earlier than Sb,
and these galaxies must have formed early ($z_f\gg3$) to provide large
enough ages for their ultraviolet luminosities to match the observed
redshift evolution and falloff at $z\sim1$. Thus, in this case we should
see a number density of galaxies in the $z=1.75\to 3$ range which is
comparable to the present-day number density ($2.2\times\ten{-3}\ h_{65}^3$
Mpc$^{-3}$ for $L > 0.25$ \lstar\ and $1.1\times\ten{-3}\ h_{65}^3$
Mpc$^{-3}$ for $L > 0.5$ \lstar).  Even if all the unidentified objects are
allocated to this redshift range there are only 10 objects, corresponding
to a number density of $4.3\times\ten{-4}\pm1.5\times\ten{-4}\ h_{65}^3$
Mpc$^{-3}$.  For \qnought=0.02 the number densities are similar, but the
problem can be avoided because the formation can begin in this redshift
range, so that number conservation is not required.  In this case only a
fraction of the disks can have begun to form at $z>2$ with the remainder
starting up at lower redshifts.

In any given redshift interval we can compute the total universal mass
density formation rate, $\dot\rho$, in observed objects, which corresponds 
to a lower limit on $\dot\rho$.  Correcting for the incompleteness is not
straightforward since it depends on the present-day type mix as a function 
of mass and the age at a given redshift.  However, including only the 
observed objects and without correcting for faint end incompleteness, we 
find mass formation rates of \mdot\ $= (1.4\times\ten{-2}, 1.7\times\ten{-2},
1.0\times\ten{-2}, 2.8\times\ten{-3})\ h_{65}^{-2}\ \msun$ Mpc$^{-3}$
yr$^{-1}$ for $z$ = (0.2, 0.6, 1.2, 2.4) and \qnought=0.02, though these
numbers are not directly comparable because of the varying cutoff in $\dot M$
so that formation rates at high redshift are relatively underestimated.  For
objects with $\dot M > 10 \msun$ yr$^{-1}$, where all objects would be
detected at all of these redshifts, the average formation rates are (0.0,
0.0, $7.1\times\ten{-3}, 2.8\times\ten{-3})\ h_ {65}^{-2}\ \msun$ Mpc$^{-3}$
yr$^{-1}$.  The overall rates are sufficient to form the present-day mass, so
that the open models do appear to provide a fully self-consistent description
with the formation peaking at $z\sim1$ at least for the objects with higher
star formation rates.

\section{Conclusion}\label{sec:conclude}

We conclude that in a \qnought=0.02 universe we can understand much of the
recent history of galaxy formation in terms of the pure luminosity
evolution of various spiral galaxy types, provided only that Sc galaxies
are at least a magnitude fainter in \itk\ than the average of galaxies
earlier than Sb, with Sds yet fainter.  Models with \qnought=0.5 predict
too many bright objects at higher redshifts.  The conclusion that this
works well for \qnought=0.02 geometries but not for flat geometries is
intimately connected with pure luminosity evolution modelling of the
number counts and redshift distributions (e.g., \markcite{met96}Metcalfe
et al.\ 1996), where \qnought=0.5 models tend to produce elbows in the
counts and fail to reproduce the faint-end counts.  The reason for this is
that in the flat models the early stages of the spirals have near-constant
and relatively bright magnitudes at all high redshifts
(\markcite{cow88}Cowie 1988) --- the same problem we encounter here with a
radically different approach.  The \qnought=0.02 models do not suffer from
this problem and have much more latitude in available time and volume to
resolve the high number count problem (\markcite{lil91}Lilly, Cowie, \&
Gardner 1991), and also do achieve broad consistency
(\markcite{met96}Metcalfe et al.\ 1996) with optical and IR number counts
and redshift distributions in this geometry.

\acknowledgments
The authors are extremely grateful to St{\'{e}}phane Charlot for comments
and help with the Bruzual-Charlot models, to Richard Ellis for
providing the Autofib data, and to Josh Barnes 
Arif Babul, Harry Ferguson, and
Mike Fall for useful comments on an earlier draft.  This
work was supported by the State of Hawaii and STScI grant GO-5922.01-94A.

\newpage

\begin{figure}
\vbox to 3.00in{\rule{0pt}{3.00in}}
\includegraphics{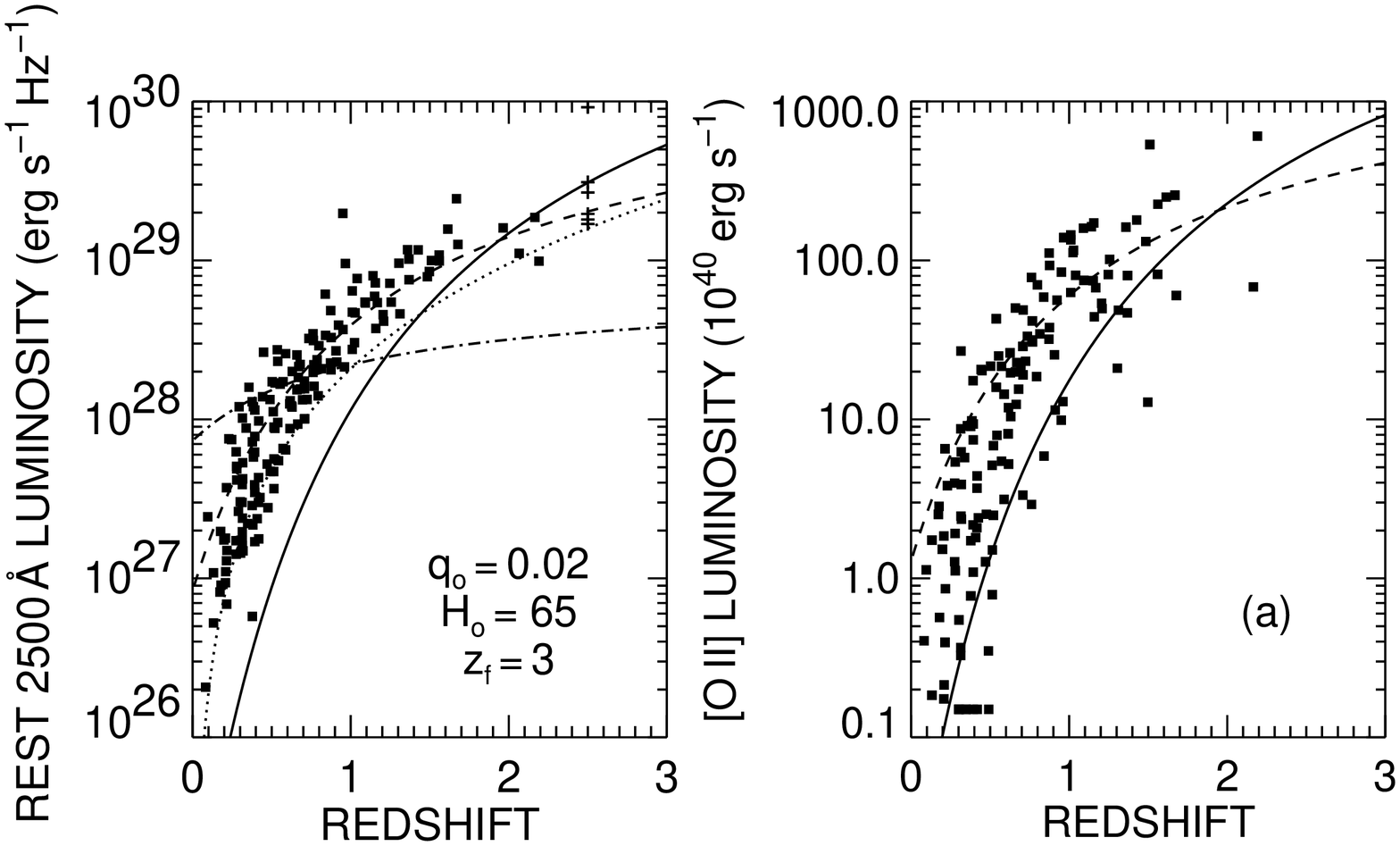}
\refstepcounter{figure}               
\label{fig:anonymous}
\end{figure}
\end{document}